\documentclass[journal]{IEEEtran}
\usepackage{xcolor,soul,framed} 
\usepackage{physics}
\usepackage{url}
\usepackage[utf8]{inputenc}
\usepackage{color}
\usepackage{graphicx}
\usepackage{graphics} 
\usepackage{epsfig} 
\usepackage{mathptmx} 
\usepackage{times} 
\usepackage{amsmath} 
\usepackage{amssymb}  
\usepackage{float} 
\usepackage{subfigure}
\usepackage{multirow}
\usepackage{pifont }
\usepackage{enumerate}
\usepackage{amsfonts}
\usepackage{cite}
\begin{document}
\bstctlcite{IEEEexample:BSTcontrol}
    \title{Towards 3D Metric GPR Imaging Based on DNN Noise Removal and Dielectric Estimation}

\author{Jinglun Feng$^{1}$, Liang Yang$^{1}$, Jizhong Xiao$^{1*}$~\IEEEmembership{Senior Member,~IEEE}
\thanks{$^{1}$The CCNY Robotics Lab, Electrical Engineering Department,
The City College of New York, New York, USA
        email(jfeng1, lyang1, jxiao@ccny.cuny.edu, $^{*}$Corresponding author.)
}
\thanks{Financial support for this study was provided by NSF grant IIP-1915721, and by the U.S. Department of Transportation, Office of the Assistant Secretary for Research and Technology (USDOT/OST-R) under Grant No. 69A3551747126 through INSPIRE University Transportation Center (http://inspire-utc.mst.edu) at Missouri University of Science and Technology. J Xiao has significant financial interest in InnovBot LLC, a company involved in R\&D and commercialization of the technology.  }
}  

\maketitle

\begin{abstract}
Ground Penetrating Radar (GPR) is one of the most important non-destructive evaluation (NDE) devices to detect subsurface objects (i.e., rebars, utility pipes) and reveal the underground scene. The two biggest challenges in GPR-based inspection are the GPR data collection and subsurface target imaging. To address these challenges, we propose a robotic solution that automates the GPR data collection process with a free motion pattern. It facilitates the 3D metric GPR imaging by tagging the pose information with GPR measurement in real time. We also introduce a deep neural network (DNN) based GPR data analysis method which includes a noise removal segmentation module to clear the noise in GPR raw data and a DielectricNet to estimate the dielectric value of subsurface media in each GPR B-scan data. We use both the field and synthetic data to verify the proposed method. Experimental results demonstrate that our proposed method can achieve better performance and faster processing speed in GPR data collection and 3D GPR imaging than other methods.

\end{abstract}

\begin{IEEEkeywords}
Back-projection (BP) algorithm, ground penetrating radar (GPR), deep neural network (DNN), none-destructive evaluation (NDE).
\end{IEEEkeywords}

\IEEEpeerreviewmaketitle

\section{Introduction}
\label{section:introduction}
\IEEEPARstart{G}{round} Penetrating Radar is widely used in non-destructive evaluation/testing (NDE/NDT), field archaeology investigation, infrastructure inspection, and measurements. GPR works by sending a pulse of polarized high-frequency electromagnetic (EM) wave into the subsurface media. EM wave attenuates as it travels in media and reflects when it encounters a material change. GPR antenna would thus record the strength and traveled time of each reflected pulse. The received signal is called an \emph{A-scan}. When GPR antenna surveys over a subsurface object, it produces a series of A-scans at different positions, and the ensemble of A-scans forms a \emph{B-scan}. The B-scan contains the hyperbolic feature which would indicate the location of the target.

However, there are two major challenges to reveal underground objects using GPR. First, it is time-consuming and tedious to manually scan a large area for detailed mapping since current commercial GPR devices can only move forward to trigger survey wheel encoder but cannot make turns. It causes the GPR data to be collected along X-Y grid lines, making the data collection vary from person to person and inconsistent from time to time. Second, the current GPR imaging techniques are easily affected by several constraints, i.e., \emph{background data noise}, \emph{permittivity of the surrounding media}, etc. \cite{pe2016automated, thitimakorn2016subsurface}. Thus, it is very important to automate the GPR data collection procedure, tag accurate pose information with each GPR sample, and propose an efficient metric GPR imaging method to allow non-professional people to understand the result of reconstructed subsurface objects.

To reach this purpose, many recent researches have devoted to the GPR imaging techniques, among which the back-projection (BP) algorithm is the most significant and commonly used GPR imaging method in NDT industry \cite{demirci2012ground}. Back projection is based on diffraction stacking that determines the location of the targets. On the other hand, the current practice of GPR data collection still requires human inspectors to pre-mark a grid map on the ground, and push the GPR cart along the straight lines \cite{xie121back}; or mount the GPR device on a track to make the measurements \cite{demirci2012ground}. Due to the above limitations, the current GPR data collection methods lack the motion freedom and the accurate metric positioning is difficult to obtain. 

GPR scanning data fused with metric information are widely researched in recent years as well. M. Pereira \cite{pereira2018new} introduced a new GPR system integration with Augmented Reality (AR) based positioning available on Google smartphones. However, the limitation in this work is that the positioning system (i.e., Google project Tango) provided by the Google is out of service now. GPR users cannot further obtain the benefits from the system they proposed. Similarly, H. Li \cite{li2019toward} introduced an automatic subsurface pipeline mapping and 3D reconstruction by fusing a GPR and a Camera. They proposed a robotic underground pipeline mapping model to facilitate GPR-based 3D reconstruction. Their major contribution is that they used the visual Simultaneous Localization and Mapping (V-SLAM), J-linkage and maximum likelihood method to estimate the radii and locations of all pipelines. However, the authors still need to pre-mark the grid map on the surface and manually push the GPR cart along the grid lines to take GPR measurements.

In this paper, we propose a novel 3D GPR imaging method that consists of three modules. A \emph{robotics data collection} module to provide a free motion pattern for GPR scanning and tag metric positioning information with GPR measurements; a \emph{background noise removal} module to clean the noisy data in the GPR B-scan images; and a \emph{DielectricNet} to estimate the dielectric property of subsurface media.

\section{Methodology}
\label{section:methodology}
\begin{figure}[H]
    \centering
    \includegraphics[width=0.48\textwidth]{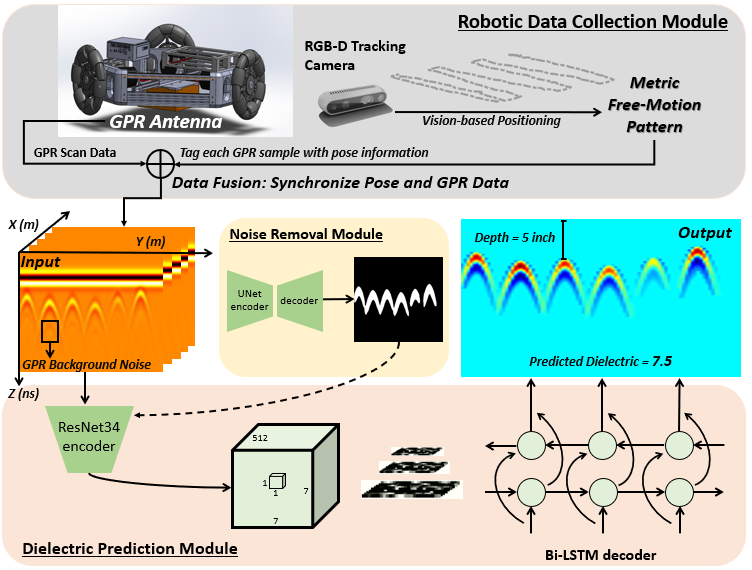}
    \caption{The system architecture consists of a robotic GPR data collection module and a DNN-based GPR data analysis software which consists of noise removal module and dielectric prediction module. }
    \label{fig:schmatic}
\end{figure}

The system architecture is illustrated in Fig.\ref{fig:schmatic}, which consists of the robotic data collection module and the GPR data analysis software. The Omni-directional robot carries a GPR antenna at the bottom of the chassis to detect and map underground objects and an RGB-D tracking camera to obtain the accurate 6-DOF pose \cite{feng2020gpr1} in real time. By tagging the GPR measurements with accurate position information in a synchronized way, it enables the robot to scan the ground surface in arbitrary and irregular trajectories and facilitates high-resolution 3D GPR imaging. The robot is able to perform a non-linear free motion that will revolutionize how GPR data is collected, interpreted, and visualized. The GPR data analysis software includes the DNN-based GPR noise removal module and the dielectric prediction module. For each input B-scan image, the noise removal module directly decodes the input image into hyperbolic features. The dielectric prediction module takes the segmentation masks from the noise removal module and pools the dielectric information from a Convolutional Recurrent Neural Network (CRNN).

\subsection{GPR 3D Metric Imaging with Free Motion Pattern}
\begin{figure}[H]
\vspace{-0.3cm}
    \centering
    \includegraphics[width=0.3\textwidth]{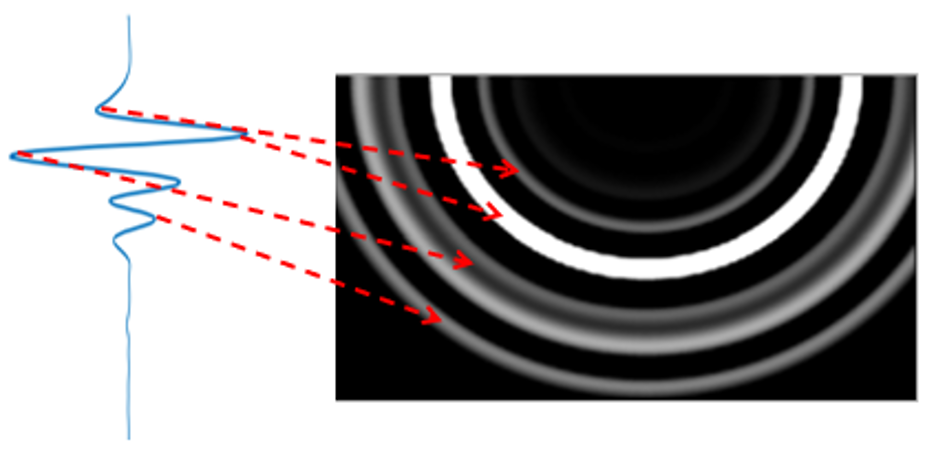}
    \caption{ BP algorithm converts the A-scan raw data into a set of semi-spheres.}
    \label{fig:bp_processing}
\end{figure}

To create a 3D GPR imaging, the BP algorithm is implemented in this section. The essence of the A-scan represents the amplitude of EM energy, while the BP is a process of aggregation that converts different amplitude of energy into a \emph{semi-sphere} format at different time. As illustrated in Fig.\ref{fig:bp_processing}, the brighter semi-sphere indicates the higher amplitude part in A-scan. Furthermore, the radius of each semi-sphere in BP image indicates the depth between the ground and the object, which is depicted by Equ.\ref{equ:bp}\cite{li2019toward}: 
\begin{align}
    \setlength{\abovecaptionskip}{0.cm}
    {\forall}A{_q^k} \in B{_k},\ (x-a){^2} + (y-b){^2} = (a{_t}*t){^2}, \ y<0
    \label{equ:bp}
\end{align} 

where $a,b$ represents the specific position of each A-scan measurement in a concrete slab. $A{_q^k} = \{a{_t}|t=1,...,n_{q}\}$ represents the $q$-th A-scan measurement in $k$-th B-scan data, while $t$ and $a{_t}$ indicate the traveling time and amplitude of A-scan signal respectively, then $n{_q}$ means the total samples in a A-scan measurements. Meanwhile, we also have $B{_k} = \{A{_q^k}|q=1,...,n{_k}\}$ that represents the $k$-th B-scan consisting of $n{_k}$ A-scans. By implementing the BP algorithm on a B-scan data, the intersection of multiple back-projected A-scan signals with the highest energy level is the possible target location.
\textbf{Free Motion Based Data Collection: }
The current practice for GPR data collection requires a human inspector to mark the grid map, push the GPR device, carefully follow the straight lines in X-Y directions, and use the wheel encoder to trigger GPR sampling. These GPR measurements are collected at equal spacing and stored slice by slice along X or Y directions. The current GPR scanning method heavily limits the freedom of motion and manually records these linear motion trajectories for GPR data post-processing. It is time-consuming and tedious to mark the grid map and perform the GPR date collection covering a large area. 
We propose an automatic GPR data collection and 3D metric GPR imaging method that combines robot control and vision-based positioning with GPR signal processing to locate and map subsurface targets. Instead of relying on a linear wheel encoder, we use an RGB-D tracking camera to estimate the position and orientation in real time \cite{yang2020rgb,feng2020gpr}. As shown in Fig.\ref{fig:schmatic}, we used an Omni-directional robot carrying both a GPR antenna and RGB-D camera that can move in any direction without spinning and thus provide a non-linear motion for the GPR scanning \cite{feng2020gpr1}. 

\begin{figure}[H]
\vspace{-0.2cm}
    \centering
    \includegraphics[width=0.45\textwidth]{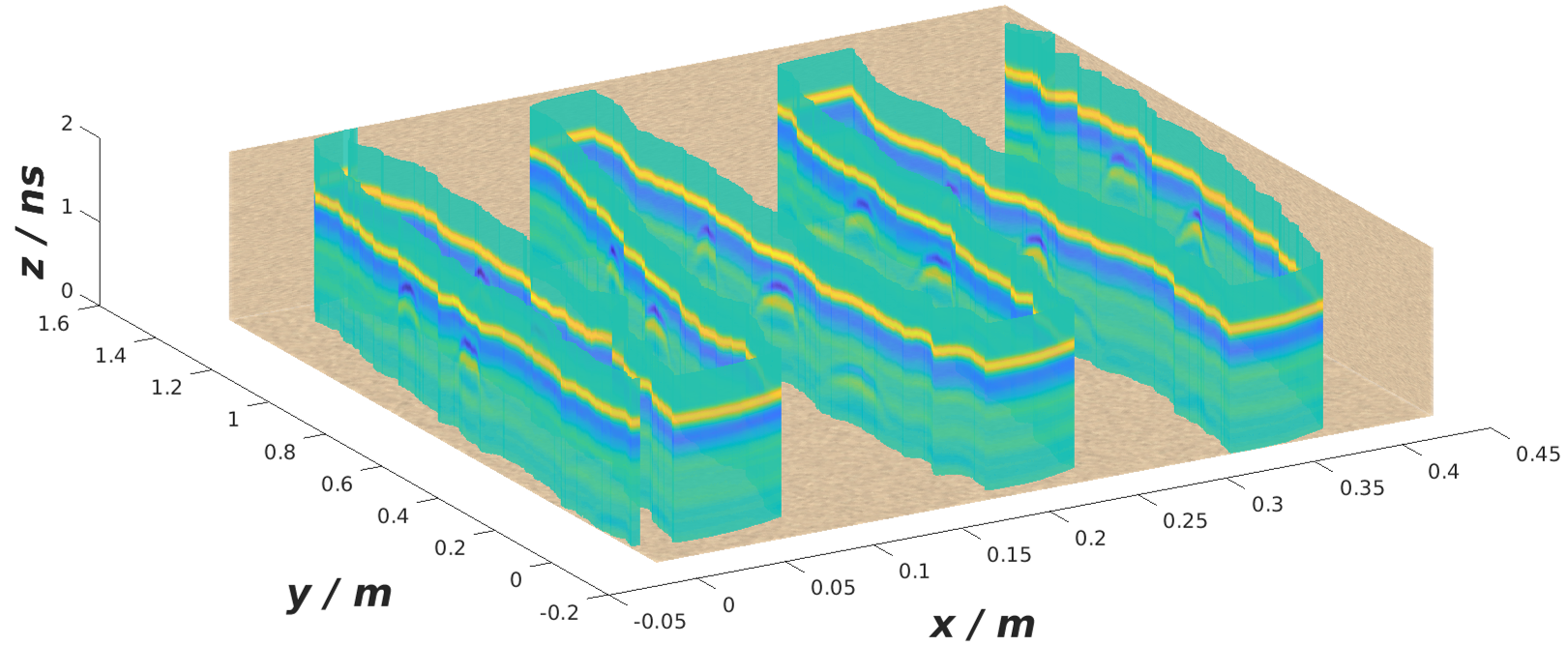}
    \caption{The B-scan profile tagged with metric positioning information.}
    \label{fig:metric}
\end{figure}

By taking advantage of this non-linear free-motion pattern, the GPR data is automatically and continuously collected and tagged with position and orientation [X, Y, Theta] at each measurement. Fig.\ref{fig:metric} illustrates how GPR B-scan data is collected along a zig-zag pattern, and it doesn't require the intervention of the human inspector.  

\subsection{DNN-based GPR Data Preprocessing}

Due to the GPR background noise and the uncertainty of the material dielectric, it is almost impossible to reconstruct the underground objects with an accurate metric. To resolve this issue, we propose two models, which are the GPR object segmentation model and the dielectric prediction model, to facilitate accurate subsurface object reconstruction and mapping. The DNN-based GPR data analytics are illustrated in Fig.\ref{fig:schmatic}, where we use a U-Net \cite{ronneberger2015u} to perform subsurface object segmentation and a DielectricNet to predict the dielectric of the subsurface material.

\subsubsection{GPR Background Noise Removal}
\label{subsection:noise_remove}

GPR works by sending an EM wave into the ground. EM wave attenuates as it travels in the medium and reflects when it encounters a change in material dielectric. GPR antenna would thus record the strength of each reflection as well as the time it takes to return to the receiver. Therefore, when GPR surveys over an underground object, some of the GPR energy pulse reflects when they hit the subsurface object, some energy continues to penetrate through the object until they become too weak to get back. However, those weak reflections would also generate hyperbolic features in the B-scan image, which could affect the GPR migration result of the targets. We call those weak responses as \textit{GPR Background Noise} (see in Fig.\ref{fig:schmatic}). Hence, it is important for us not only to focus on enhancing and sharping B-scan features but also removing those background noise.

Specifically, we firstly get the hyperbolic masks from the input raw B-scan images by taking advantage of a segmentation model (i.e., UNet). Then, we filter the raw B-scan data with mask B-scan features so that the filtered data only keeps the strongest target reflection signal.

\subsubsection{GPR Targets Depth Prediction}
\label{subsection:dielectricnet}

The dielectric constant is crucial for the GPR migration process since it determines the depth of each target shown in a B-scan image. As shown in Equ.\ref{equ:dielectric}, the EM wave which is emitted from the GPR antenna is supposed to be the speed of light in a vacuum. However, the dielectric ${\mathcal{D}}$ of subsurface materials would make a difference to the traveling speed, which further impacts the depth prediction of the target.

    \vspace{-0.3cm}    
    \begin{align}
    v &= \frac{C} {\sqrt{\mathcal{D}}} \\
    {D}_{tar} &= \frac{{T}_{tr} * v}{2}
    \label{equ:dielectric}
    \end{align}


where ${D}_{tar}$ denotes the depth of subsurface targets while ${T}_{tr}$ is the two-way travel time of GPR EM signal.  

Thus, for any underground object, if the dielectric of the surrounding material is high, the B-scan hyperbolic feature would be shallower than the same target with a lower value of the environment dielectric. On account of this property and inspired by \cite{7801919}, we introduced a CRNN model in this letter (i.e., DielectricNet) to estimate the dielectric, $d_{t(i)}, i \in \{0,1,...\}$ for each target.

Given a GPR scan in media with an unknown dielectric, an informative and discriminative feature representation plays a significant role in dielectric estimation. The feature representation should preserve the dielectric property of the underground media. Moreover, it should consider the spatial distribution of all hyperbolic features to enable better dielectric prediction. We take advantage of Recurrent Neural Network (RNN), which is able to take a sequence of signals, either spatially continuous or temporally continuous, to estimate the dielectric. The $ResNet34$ \cite{he2016deep} is implemented as the encoder and an RNN model as the decoder, where the GPR images are fed into a series of convolutional layer to obtain the feature map that contains encoded dielectric information \cite{rong2017unambiguous}.

In our implementation, the input noise-free B-scan images are resized to $224 \times 224$ while the $ResNet34$ architecture is changed to work in a fully convolutional manner by converting \textit{fc6} to fully convolutional layers. Then, a {bidirectional long short-term memory} (\textit{Bi-LSTM}) decoder extracts dielectric property from the CNN encoded feature descriptors, where it sequentially outputs their corresponding confidence scores for the predicted dielectric value. This design helps the model to reason the dielectric property.

\textbf{Loss Design:} Once the dielectric property is predicted, we optimize the model by using a Connectionist Temporal Categorical (CTC) loss.

\begin{equation}
    \centering
    \sum_{i}P( \mathcal{D}\mid \mathbf x) = \sum_{i}P(\mathcal{D} \mid \pi_{i})P(\pi_{i} \mid x)
    \label{equ:faster_rcnn}
\end{equation}

where ${\mathcal{D}}$ is the ground truth of the dielectric property, $\mathbf x$ is the input sequence that comes from the RNN model, and $\pi_{i}$ denotes each possible value of the dielectric. The depth of the subsurface target can be further calculated according to Equ.\ref{equ:dielectric}.

\section{Experimental Study}
To demonstrate the effectiveness of the proposed 3D metric GPR imaging method, we conducted both simulation and experimental study. The synthetic data were generated by gprMax \cite{warren2016gprmax} while the field data were collected on a concrete slab at CCNY Robotics Lab Testing Pit. We also compare the effectiveness between our method and other approaches, and discuss the effectiveness of the DielectricNet model.

To create the dataset for experimental purposes, we have collected two kinds of data: 1) the field data collected on CCNY Testing Pit, which contains $120$ B-scan data; 2) synthetic data we proposed in \cite{Feng_2021_WACV}, and it contains $628$ B-scan data. Note that all the B-scan images are taken as the input for noise cancellation and dielectric prediction task and then used to verify the proposed method.

\begin{figure*}[htbp]
\centering
    \vspace{-0.5cm}
    \subfigure[The ground truth of the synthetic slab generated by gprMax.]{
        \includegraphics[width=0.44\textwidth]{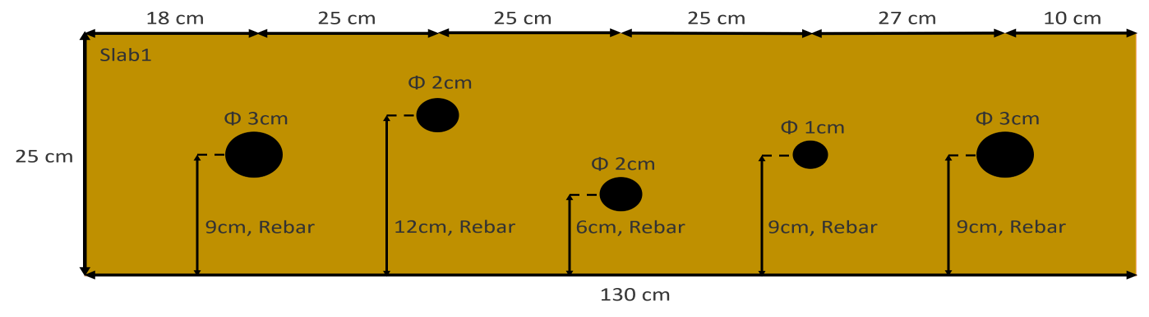}
    }
    \subfigure[The ground truth of the field CCNY Robotics Lab concrete slab.]{
    	\includegraphics[width=0.44\textwidth]{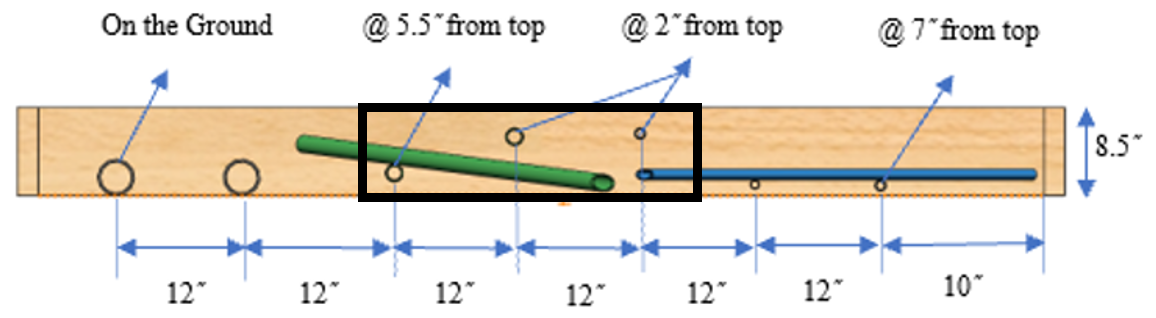}
    
    }
    \quad
    \vspace{-0.3cm} 
    \subfigure[The raw B-scan of the synthetic slab.]{
        \includegraphics[width=0.44\textwidth]{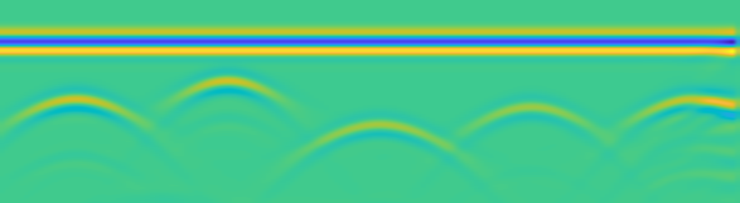}
    }
    \subfigure[The raw B-scan of the field slab.]{
        \includegraphics[width=0.44\textwidth]{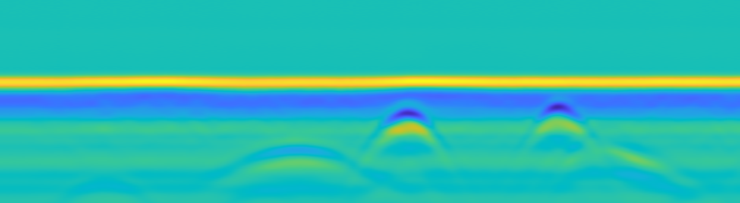}
    }
    \quad
    \vspace{-0.1cm} 
    \subfigure[Filtered B-scan of the synthetic slab.]{
        \includegraphics[width=0.44\textwidth]{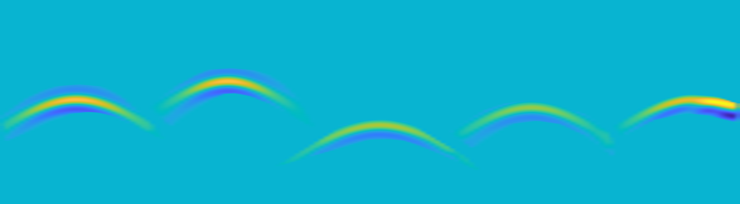}
    }
    \subfigure[Filtered B-scan of the field slab.]{
        \includegraphics[width=0.44\textwidth]{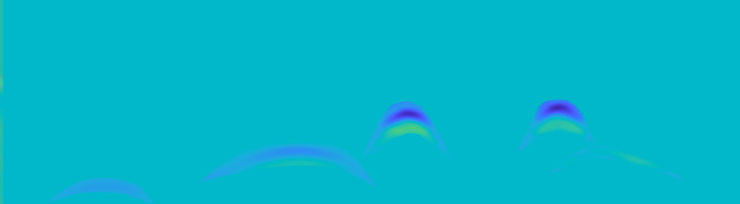}
    }
    \quad
    \vspace{-0.1cm} 
    \subfigure[The migration result with the synthetic GPR data by proposed 3D metric method.]{
        \includegraphics[width=0.45\textwidth]{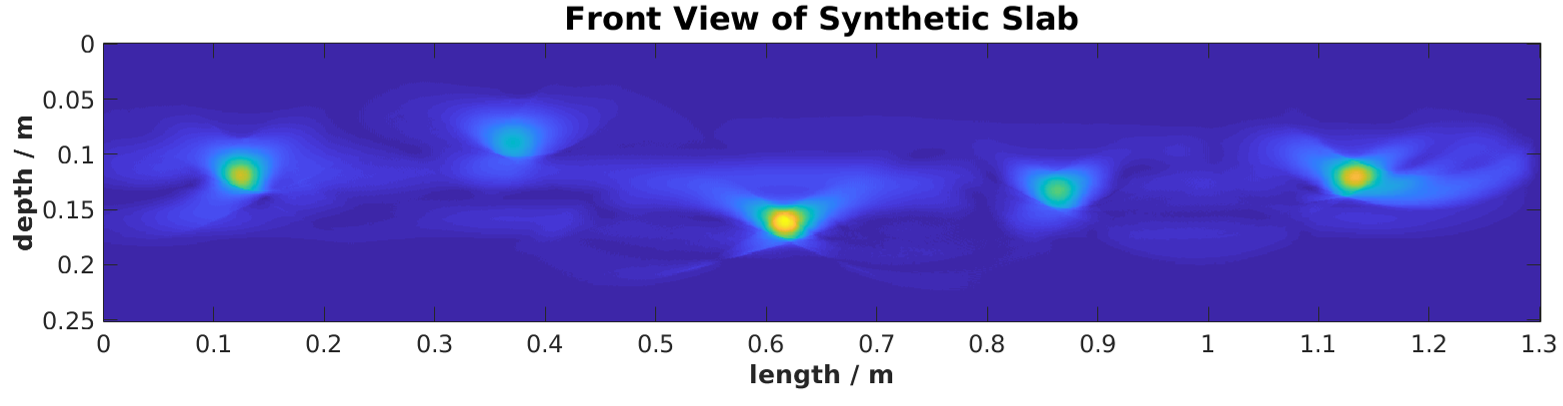}
    }
    \subfigure[The migration result with the field GPR data by proposed 3D metric method.]{
        \includegraphics[width=0.45\textwidth]{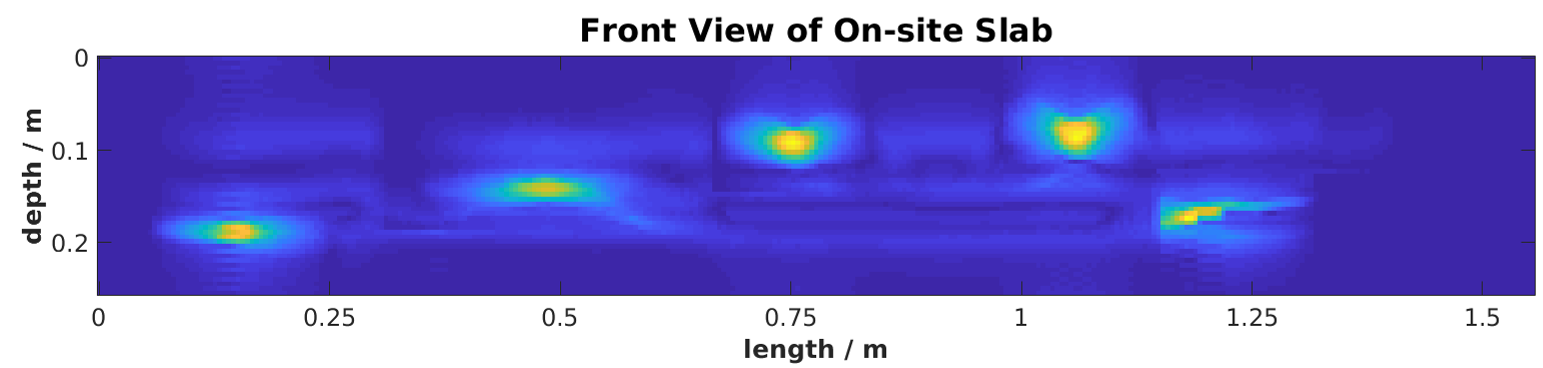}
    }
    \quad
    \vspace{-0.1cm} 
    \subfigure[The migration result with the synthetic GPR data by the conventional method\cite{xie121back}.]{
        \includegraphics[width=0.45\textwidth]{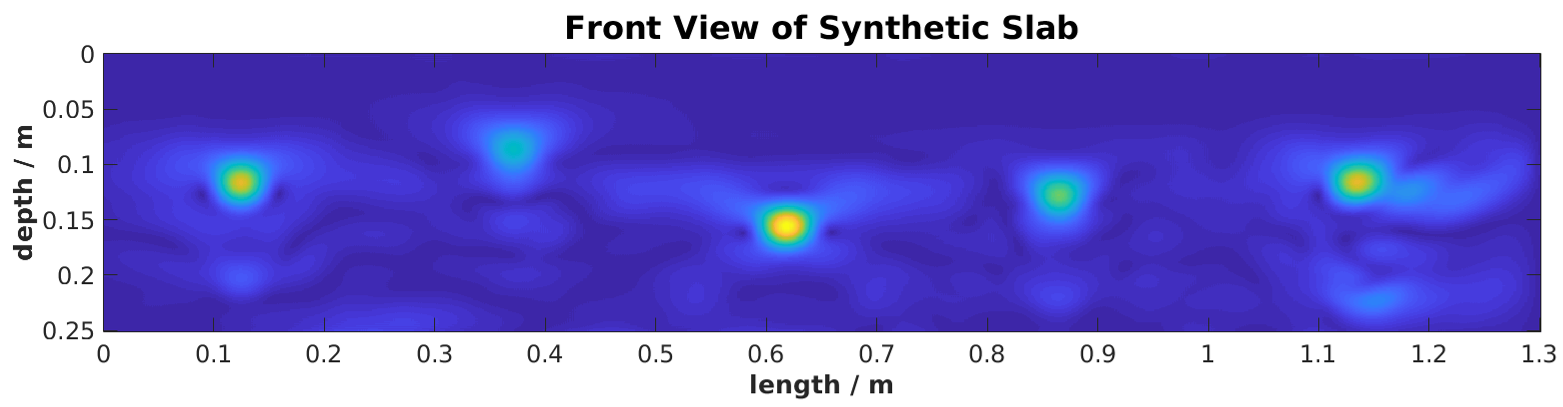}
    }
    \subfigure[The migration result with the field GPR data by the conventional method\cite{xie121back}.]{
    	\includegraphics[width=0.45\textwidth]{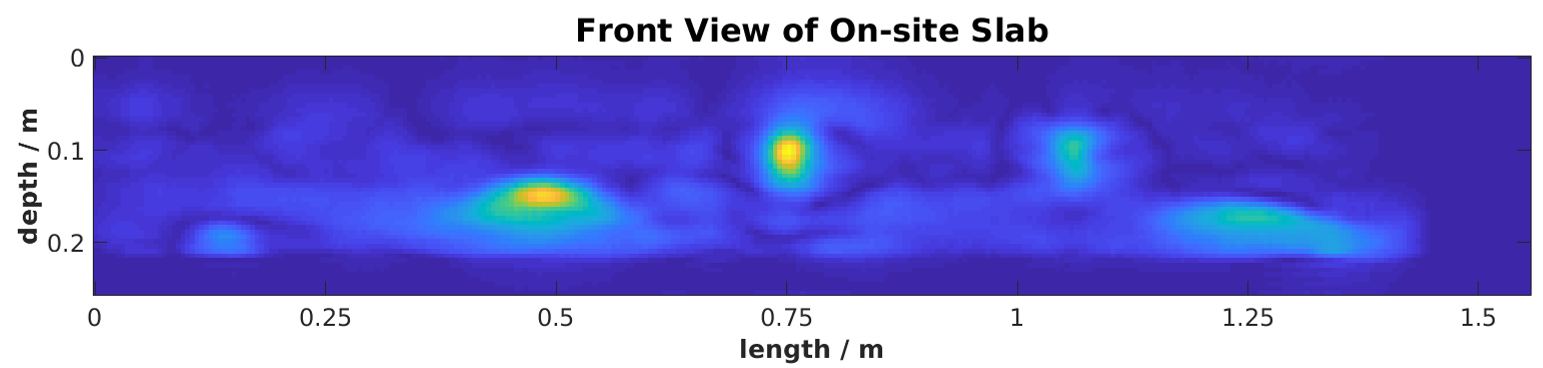}
    
    }
    \quad
    \subfigure[The migration result with the synthetic GPR data by the method proposed in \cite{feng2020gpr}.]{
        \includegraphics[width=0.45\textwidth]{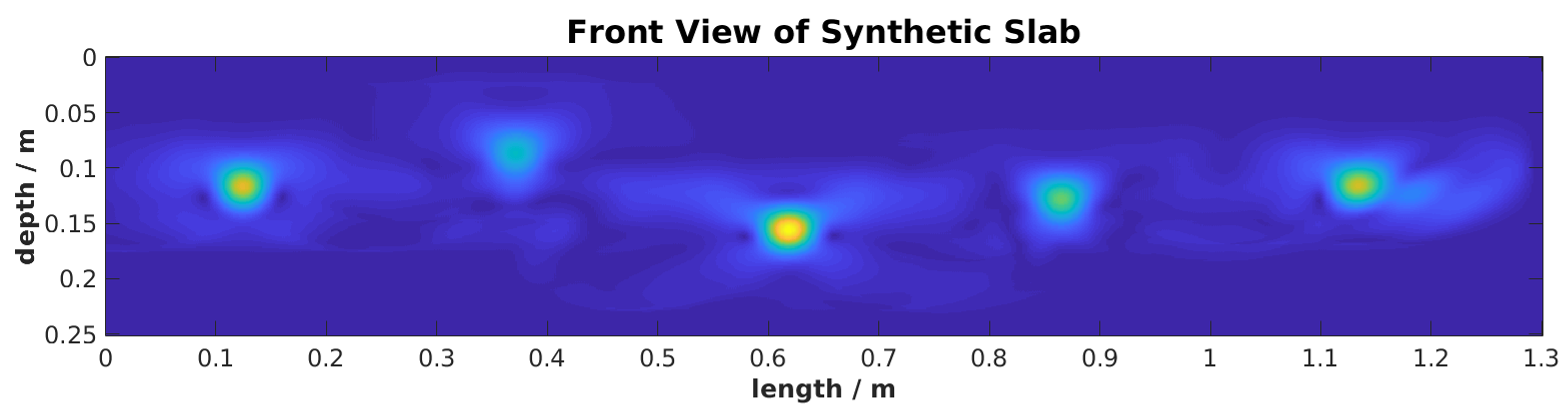}
    }
    \subfigure[The migration result with the field GPR data by the method proposed in \cite{feng2020gpr}.]{
        \includegraphics[width=0.45\textwidth]{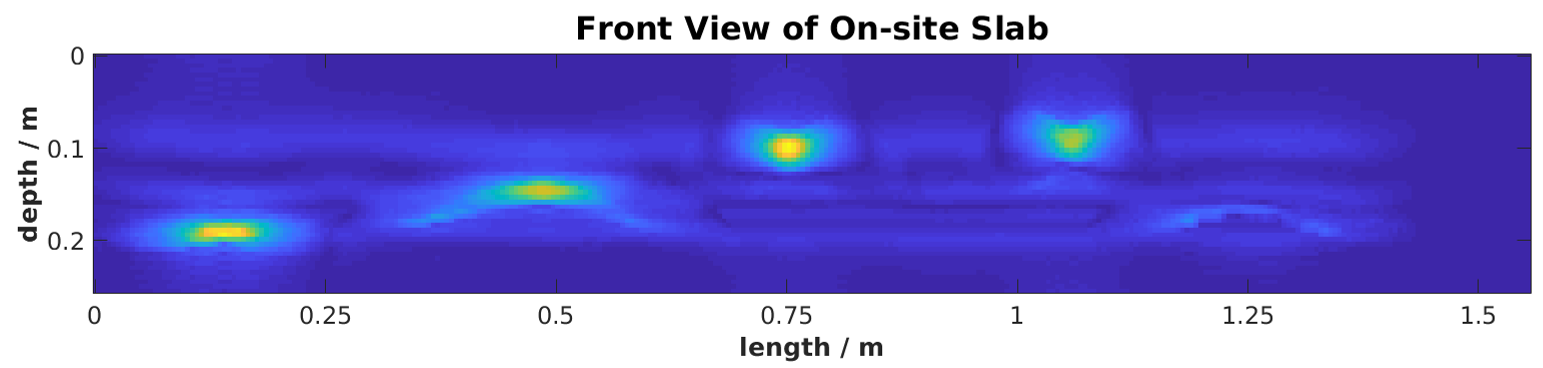}
    }

    \caption{Effectiveness and noise comparison of proposed method on both synthetic and field GPR data, all the 3D GPR imaging results are shown with a front view. Note that in subfigure $(b)$, the black bounding box demonstrates the field data collection area.}
    \label{fig.effectiveness}
\end{figure*}

\subsection{Comparison Study}
As illustrate in Fig.\ref{fig.effectiveness}, both synthetic and field GPR data have been used to validate the effectiveness of our method. Specifically, Fig.\ref{fig.effectiveness}$(a)$ shows a synthetic slab which is $1.3m$ meters long and $0.25$ meters deep. In addition, Fig.\ref{fig.effectiveness}$(b)$ shows the ground truth of the field concrete slab, note that the black bounding box showing in $(b)$ demonstrates the field GPR data collection area, which is $1.55$ meters long and $0.2$ meters deep. Fig.\ref{fig.effectiveness}$(c)$ to $(f)$ shows the raw B-scan data and noise removed B-scan data respectively. Fig.\ref{fig.effectiveness}$(g)$ and $(h)$ shows the proposed 3D metric GPR imaging results.



To further validate the proposed method, we compare our approach with the conventional BP method \cite{xie121back} and our previous DNN-based migration method proposed in \cite{feng2020gpr}, where different DNN architecture was used to remove the noise in B-scan data. 

Fig.\ref{fig.effectiveness}$(g)$-$(l)$ demonstrate GPR imaging results generated by the above methods respectively. Specifically, $(i)$ and $(j)$ show that the conventional BP results contain too much noisy data while in subfigure $(k)$ and $(l)$, the noise is removed and the result is sharper using our old DNN-based migration method. The best 3D GPR imaging results are shown in subfigure $(g)$ and $(h)$ using our new 3D metric migration method.

Table.\ref{table:crnn} provides quantitative evaluation of these three methods in terms of the GPR image noise level and similarity effectiveness, which are evaluated with four different metrics: \textit{Euclidean Distance}, \textit{Mean Square Error}, \textit{Signal-to-Noise-Ratio} and \textit{Structural Similarity Index} on both synthetic dataset and field dataset. For Euclidean Distance and Mean Square Error, the lower the value, the better performance is. For Signal-to-Noise-Ratio and Structural Similarity Index, the larger the value, the better effectiveness it represents. Our new DNN-based method outperformed other methods.

\begin{table}[!th]
\caption{Effectiveness Comparison between proposed method and others with multiple metrics}
\label{table:weight}
\begin{center}
\begin{tabular}{|c|c|c|c|c|c|c|c|c|c|c|}
\hline
\hline
&\multicolumn{2}{|c|}{E\_distance} &\multicolumn{2}{|c|}{MSE} &\multicolumn{2}{|c|}{SNR} &\multicolumn{2}{|c|}{SSMI}\\ 
\hline
\hline
Ours w/ synthetic dataset & \multicolumn{2}{|c|}{\textbf{2.23e+04}}&\multicolumn{2}{|c|}{\textbf{284.774}} &  \multicolumn{2}{|c|}{\textbf{16.982}} & \multicolumn{2}{|c|}{\textbf{0.987}}\\
\cite{feng2020gpr} w/ synthetic dataset & \multicolumn{2}{|c|}{4.05e+04} &\multicolumn{2}{|c|}{351.455} &  \multicolumn{2}{|c|}{16.068} & \multicolumn{2}{|c|}{0.976} \\
\cite{xie121back} w/ synthetic dataset & \multicolumn{2}{|c|}{9.08e+04}&\multicolumn{2}{|c|}{453.541} &  \multicolumn{2}{|c|}{14.961} & \multicolumn{2}{|c|}{0.983} \\
\hline
\hline
Ours w/ field dataset & \multicolumn{2}{|c|}{\textbf{1.39e+03}} &\multicolumn{2}{|c|}{\textbf{397.384}} &  \multicolumn{2}{|c|}{\textbf{15.810}} & \multicolumn{2}{|c|}{\textbf{0.980}} \\
\cite{feng2020gpr} w/ field dataset & \multicolumn{2}{|c|}{1.92e+03} &\multicolumn{2}{|c|}{485.132} &  \multicolumn{2}{|c|}{14.943} & \multicolumn{2}{|c|}{0.971} \\
\cite{xie121back} w/ field dataset & \multicolumn{2}{|c|}{3.40e+03} &\multicolumn{2}{|c|}{983.156} &  \multicolumn{2}{|c|}{11.876} & \multicolumn{2}{|c|}{0.938} \\
\hline
\end{tabular}
\end{center}
\end{table}

The reason is that a segmentation model could capture the shape of the hyperbolic feature, which contains all the useful information in B-scan data. In contrast, the information in the bounding box region generated by the detection-oriented model would contain some noise data besides the useful B-scan data.

\subsection{Why CRNN model is used for dielectric prediction?}
The CRNN model has widely used for the \textit{scene text recognition} tasks in computer vision community \cite{7801919}. The input to the CRNN model is the image that contains text, while the output is the text sequence extracted from the image. This end-to-end method is proven to be effective and inspired us to consider that the hyperbolic features contained in GPR B-scan images could also be analyzed in a similar way by the CRNN model. The hyperbolic features are normally determined by several factors including the dielectric property of surrounding media, the dimension, material, and depth of the buried utility pipelines. As long as the dataset contains enough objects with different combinations of the above properties, the CRNN model shall be able to extract the dielectric value of the surrounding media with the various values of other factors. In our dataset, the dimension, material, and depth of the targets are illustrated in Fig.\ref{fig.effectiveness} (a) and (b). Since the dielectric value and depth of hyperbolic features are correlated, we can use a CRNN model to encode this property as an abstract sequence, and further decode and estimate this abstract sequence as a specific dielectric value. Specifically, the CNN encoder we implement is \textit{ResNet34} and the RNN decoder is \textit{Bi-LSTM}. In addition, different methods, which are a CRNN model with \textit{VGG16} encoder and \textit{Bi-LSTM} decoder and a single \textit{RseNet34} network\cite{feng2020gpr} respectively, are illustrated by comparing their prediction accuracy and running time, since the loss function in each method is not the same. As shown in Table \ref{table:crnn}, our CRNN model achieves the highest dielectric prediction accuracy and at least $24.6\%$ faster in comparison with other baseline methods.

\begin{table}[!th]
\caption{Performance Comparison between proposed method and others} 
\begin{center}
\begin{tabular}{|c|c|c|c|c|c|c|c|c|}
\hline
\hline
&\multicolumn{2}{|c|}{Ours} &\multicolumn{2}{|c|}{VGG16 + Bi-LSTM} &\multicolumn{2}{|c|}{ResNet34}\\ 
\hline
\hline
Acc. & \multicolumn{2}{|c|}{\textbf{98.29}}&\multicolumn{2}{|c|}{97.56} &  \multicolumn{2}{|c|}{{98.05}} \\
Time(sec.) & \multicolumn{2}{|c|}{\textbf{0.0049}} &\multicolumn{2}{|c|}{0.0065} &  \multicolumn{2}{|c|}{{0.0075}} \\
\hline
\end{tabular}
\end{center}
\label{table:crnn}
\end{table}
\section{Conclusion}
This paper introduces a DNN-based 3D metric GPR imaging method, which can enhance the GPR imaging performance by taking the following steps. 1) Introducing the robotic GPR data collection that provides free motion pattern for GPR scanning and tags the position information with GPR measurements. 2) Using a DNN-based segmentation model to remove the background noise from raw GPR B-scan images. 3) Using DielectricNet to estimate the dielectric value of the underground environment. Experimental results show that our proposed method performs better in comparison with other approaches. Our method makes the GPR data collection significantly easier by enabling the automatic scan of the flat surface in arbitrary trajectories with minimal human intervention. It eliminates the time, hassle, and cost to laying out grid lines on flat terrain and reducing the hassle to closely follow the grid lines and the note-taken time to record the linear motion trajectory in X-Y directions. Our 3D metric GPR imaging software processes the GPR data tagged with pose information to reveal subsurface objects with sharper images and make inspection results easy to understand. 


\ifCLASSOPTIONcaptionsoff
  \newpage
\fi

\bibliographystyle{IEEEtran}
\bibliography{Bibliography}

\end{document}